# The Effect of Providing Peer Information on Evaluation for Gender Equalized and ESG Oriented Firms: An Internet Survey Experiment

Eiji Yamamura.    Seinan Gakuin University, Japan

Abstract

Internet survey experiment is conducted to examine how providing peer information of evaluation about progressive firms changed individual's evaluations. Using large sample including over 13,000 observations collected by two-step experimental surveys, I found; (1) provision of the information leads individuals to expect higher probability of rising of stocks and be more willing to buy it. (2) the effect on willingness to buy is larger than the expected probability of stock price rising, (3) The effect for woman is larger than for man. (4) individuals who prefer environment (woman's empowerment) become more willing to buy stock of pro-environment (gender-balanced) firms than others if they have the information.  (5) The effect of the peer information is larger for individuals with "warm -glow" motivation.

# I. Introduction

Firms seem to put more importance on environment, Society and Governance (Hereafter, ESG) and gender-equalization to meet requirements from society. In response to it, economic researchers increasingly have paid attention to role of firms in terms of social responsibility and sustainable society[1]. In addition to ESG, woman's role in firms become important and their impact on firm performance is also explored (e.g., Kawaguci 2007; Abern and Dittmaer, 2012; Mast and Miller 2013; Gangadharan et al., 2019).

The ESG oriented firms also experienced higher profitability, growth, and sales per employee than other firms (e.g., Wu and Shen 2013; Chan and Walter,2014; Gupta et al. 2015; Lins et al., 2017). Hence, people possibly pursue the return from the ESG or woman empowerment firms, and so buy stock of these firms. That is, it is unknown about how do people consider importance of ESG and woman empowerment when they evaluate firm. The rudimentary question is not sufficiently addressed by researchers. It is difficult to scrutinize the question using data of real world.

Hence, this study analyzes it by conducting an internet experiment which is increasing employed by researchers (e.g., Kuzemko et al., 2015; Fisman et al., 2020; Bimonte et al. 2020). Enjoying advantage of the internet survey, its sample size is over 13,000 which are gathered by the first and follow-up surveys. As is observed in existing works, the peer information enhances the pro-social behaviors. In the experiment, in the first survey,

---

[1] Information provided on media coverage on ESG engagements with local impact on companies' communities and employees is key factor to increase shareholder value and improve operating performance (Byun and Oh 2018). On the demand side, investors exhibit optimistic responses to good news about companies with higher ESG scores but pessimistic responses to bad news about companies with lower ESG scores (Chen and Yang,2020). Socially connected fund manager's decision in the market is influenced by neighbors (Pool et al, 2015).

I gathered data about subjective probability of rise of ESG firm stock price, and willingness to buy its stock. In the follow-up survey, we provide respondents the information of percentages of those who expect the stock price to rise and of buying the stock gathered in the first survey. Then, we asked them the same questions. I compare effect of the peer information on the probability of the stock price rising and on willingness to buy the stock. People put importance on the ESG activity and woman empowerment if effect on the willingness to buy the stock is larger than that on the expected stock price rising.

Existing works used field-experiment to examine effects of providing peer information and found that peer information enhanced pro-environmental behaviors. Concerning social comparison-based home energy reports (HER) in the U.S., information of comparing residents' electricity use to that of their neighbors reduces energy consumption (Allcott 2011) and the effect of repeated reports of social comparison persists even though time has passed (Allcott and Rogers,2014). These imply that the HER is a cost-effective climate policy intervention. However, a similar field experiment in Germany has a smaller effect than the U.S (Andor et al. 2020). Therefore, the policy effect varies according to norms and preferences shared by society (Yeomans and Herberich,2014). Altruistic motivation leads to invest for firm's ESG activities (Jha and Cox 2015). Investors are more likely to invest in firm based on their social preferences than their financial performance (Riedl and Smeets 2017).

Charitable donation is considered as one of pro-social behaviors. Motivation of the donation is possibly based not only on pure-altruism but also warm-glow (Andreoni 1989; 1990). Pure altruistic people are motivated solely by an interest in the welfare of the recipients of their donation. Instead of it, warm-glow people receive the positive emotional feeling from helping others. These possibly explain the pro-environmental and pro-social behaviors such as buying stock of ESG firms.

Further, gender-difference is widely observed concerning investment. male is more likely to be over-confident to have aggressive decision making in the stock market (Barber and Odean, 2001; Cueva et al. 2019). According to study the impact of gender on asset allocation recommendation, male students choose a riskier allocation than female students. In contrast, male and female finance professionals feature similar risk preferences. Social framing tends to reinforce prosocial behavior in women but not men (Espinosa and Kovářík 2015). Researchers found that women give more to charitable organizations than men do (e.g., Andreoni et al., 2003; DellaVigna et al., 2013).

So, it is valuable of examining how the effect of peer information varies according to preference, subjective value and sex. Further, Environment, Social, Governance, and Woman are obviously different notion[2]. Hence, ESG oriented firms should not be treated differently according to sub-categories. For instance, what people expect pro-environment firms to contribute is different from what they expect well-governed transparent firms. So, examinations are conducted to compare effect of the peer information between different type of ESG firms.

In this study, I found that provision of the peer information raises individual's expected probability of rising of stock price and makes them more willing to buy it. Further, the effect on the willingness to buy is larger than the probability of stock price rising.   Effects of the information on woman are larger than on man. Further, individuals who prefer pro-environmental and woman empowerment society, and warm-glow donation become likely to buy the stock if they are informed of peer information. Hence, provision of peer information is considered as "nudge" to increase their confidence to behaver based on their preferences.

---

[2] Well-governed firms tend to engage in prosocial activities (Ferrell et al, 2016).

The remainder of the paper is set out as follows. In Section II I describe the method of experimental strategy. In Section III I provide an overview of the data and the descriptive statistics. In Section IV, I explain the estimation approach, while Sections V provides the estimation results and its interpretation. Section VI summarizes our conclusions and draws out some implications for future research.

**II. Experimental design and data and**

*A. Internet experiment*

I independently collected individual-level data through internet-surveys covering all parts of Japan in 2018. Details of the sampling method are provided in the Appendix.

During 25-30th October 2018, I conducted the first survey to obtain data of socio-economic and demographic data, such as respondents' genders, ages, household incomes, job status, marital status, and number of siblings. Further, subjective evaluation about four types of firms, their willingness to buy stock of these firms. Questions and respondent's choices are exhibited in Table 1. There are two types of questions and 5 choices in each question as. To take an example of pro-environment firm;

*Do you agree that stock of the firm with female board members will rise?*

*5 (strongly agree)- 1 (strongly disagree)*

*Are you willing to buy stock of firm with female board members?*

*5 (strongly wish to buy)- 1 (no, not at all)*

Respondents are asked to answer two questions about 4 different types of firm; (1) *gender balanced firm*, which defined as firm with female board members in this study, (2) *pro-environment firm,* (3) *pro-social firm,* (4) *transparent firm.* The first type of firm is considered as gender-balanced firm. The second, third and forth firms are considered as group of the ESG firm.

Apart from two types of questions concerning stock of firm, respondents are asked about preference for woman empowerment. Evaluation about gender-balanced firm is considered to depend on the preference. The question and respondent's choices are;

> *Government should form a society in which women can demonstrate their ability and be actively involved in the work place.*
>
>   *5 (strongly agree)- 1 (strongly disagree)*

Similarly, respondents are asked about preference for environment, which is related to evaluation about pro-environmental firm. The question and respondent's choices are;

> *Government should put more emphasis on environment countermeasure..*
>
>   *5 (strongly agree)- 1 (strongly disagree).*

Generally, evaluation about ESG is possibly depend on charitable preference because people with charitable motivation are unlikely to pursue return from its stock. Hence, I asked about preference for charity. Respondents are asked to exhibit amount of tax one can pay assuming respondent gains distinctly higher income than majority of people. In order to distinguish warm-glow motivation from pure-altruism motivation, respondents are asked about the same questions on the assumption that double of amount of tax they would pay is directly redistributed to lower income people. If the pay more, they are considered to have pure-altruistic motivation rather than warm-glow. In other words, those who do not want to increase tax have warm-glow motivation because they do not consider degree of contribution to improve lower income people's economic condition. That is, warm-glow people do not change behavior and evaluation even if they can more contribute to improve economic condition of poor people. In any questions as above, there is no choice of "unknown" or "do not reply". Further, survey is completed only if respondents answer all questions.

After about two-weeks, during 26-21th November 2018, the follow-up survey has been conducted to ask the same questions about firm evaluation and willingness to buy its stock. One thing different from the first survey is they answer the question in which the following peer information is exhibited; rates of those who expect the firm's stock to rise, and also rates of willingness to buy its stock in the first survey. To take an example of gender-balanced firm;

*Based on the firs-survey, 12 percentage of respondents expect that the stock price of the firm with female board members to rise. Do you agree that stock of the firm will rise?*
    *5 (strongly agree)- 1 (strongly disagree)*

*Based on the firs-survey, 15 percentage of respondents are willing to buy stock of the firm with female board members. Are you willing to buy stock of the firm?*
    *5 (strongly wish to buy)- 1 (no, not at all)*

In Appendix, Figure A1 demonstrates one of screen where respondents see when they are asked about willingness to buy stock of gender-balanced firm. Percentage of who chose "*strongly wish to buy"* or *"wish to buy*" stock of firm with female board members is indicated by colored in red after its explanation (15%). In the similar way, in other questions, respondents can know the peer information in the first wave and then answer the question. These are defined as rate of respondents who chose "4" or "5" from choices in each question. Apart from provision of peer information, we can assume other things are considered as constant because there is only two-weeks between the first and the follow-up surveys. Therefore, change of their choice from the first to the follow-up surveys reflects influences of the degree of other people have positive view about the firm.

*B. Data*

I invited 9,300 subjects and then gathered 7,855 observations, and so response rate is 84.4%. In this study, I restricted sample of those who responded to the first and the follow-up surveys because we compare subjective evaluation and willingness to buy between before and after providing peer information. So, 1,141 respondents completed the first survey but did not participate in the follow-up one. Hence, 6714 individuals competed the first and follow-up surveys and are included in the sample for estimation. Accordingly, sample size becomes 13,428. Male and female observations are 7,048 and 6,380, respectively.

Figures 1 (a)-(d) compare subjective evaluation about each type of firm between genders in the first survey and so before knowing peer information about the evaluation. Woman's expectation to rise and willingness to buy are significantly higher than man in all figures. When it comes to comparison between expectation to rise and willingness to buy, willingness to buy is lower than expectation to rise in most cases for gender-balanced, pro-environment and pro-social firms. This indicates that individuals are less likely to have confidence in their expectation to rise.

However, in exceptional cases that are pro-environment and pro-social firm for woman, I do not find difference between expectation to rise and willingness to buy. In one interpretation, women are more confident than man about pro-environment and pro-social firms. However, women are generally less confident than man in decision making for buying stock (Barber and Odean, 2001; Cueva et al. 2019). Therefore, woman evaluate these firms not only by its return from its stock but also by its contribution to environment and society. Turning to Figure 1 (d) showing observation of Transparent firms, for both man and woman, individual's willingness to buy is higher than expectation to rise. This indicates that transparent firms are reliable and its stock price is likely to be stable. Hence, individuals with risk averse are more likely to buy its stock even if its stock price is unlikely to rise.

Figure 2 (a) compare the preference for gender equality and environment between man and woman in the first wave. This clearly show that women are more likely to prefer gender equality and environment than man, which is consistent with Figures 1 (a) and (b). Figure 2 (b) compare charitable preferences between genders in the first wave, which is standardized to compare different measure. Surprising, man prefers charity than woman.

Table 1 shows the description of variables and mean values of before and after the experiment (first and follow-up surveys). Regardless of characteristics of firms, Mean values of key variables examined, individual's expected probability of the stock price rising, and willingness to by the stock, are larger after the providing the peer information than before it.

## III. The Econometric Model

In each type of firms, I estimated the effect of the peer information on the probability of the stock price rising and on willingness to buy the stock. Both of former and latter effects are expected to be positive. Further, I calculate the value "the latter effect" over "the former effect". In the case that the value is larger than 1, people's evaluation about the ESG activity and woman empowerment leads them to buy the stock because increase of willingness to buy is larger than increase of expectation about the stock price rising.

I assume that well-governed transparent firm is different from other firms such as pro-environment, pro-social, and gender balanced firms because the transparent firms are expected to improve firm performance to raise the its stock price. So, evaluation about transparent firm is based on self-interest. Comparison between transparent firm and other firms is useful to scrutinize how people are motivated by support for environment, society, and woman involvement.

In the baseline model, the estimated function takes the following form:

$$\textit{Rise (or Buy )}_{it} = \alpha_1 \textit{ Information }_t + m_i + u_{it}, \qquad (1)$$

The dependent variable *Rise (*or *Buy)* denotes the individual *i*'s expected probability that stock price of firm *k* or individual *i*'s willingness to buy the stock on timing of survey *t*. *Information* is dummy for follow-up survey which captures giving peer information and $\alpha_1$ is its coefficient. $m_i$ is individual's dummies to capture time-invariant individual-level factors. There is only two-weeks between the first and follow-up surveys and so most of individual's factors do not change. Hence, the simple specification above enables me to examine effects of giving peer information. $u_{it}$ is error term.

As is observed in Figure 1, women are more likely to have positive view about ESG and gender-balanced firms. However, possibly Women are less confident in their stock trading than men (Barber and Odean, 2001; Cueva et al. 2019). Peer information possibly leads them to be more confident and change their behaviors (Allcott 2011; Allcott and Rogers,2014).    In compared to firms to pursue maximizing its profit, the ESG firms are not sufficiently recognized in society. Expectation about stock price rising depends on demand in the market. The situation is similar to voting behavior in beauty contest and so people expect that the firm's stock price rises not because they prefer it but because others will buy it. Accordingly, individual's preference is unlikely to be reflected in evaluation about firm. However, peer information possibly leads their preference to be reflected in evaluation.    *Information* is interacted with woman dummy to explore difference of effect of *Information* between genders. Further, *Information* is interacted with preference for woman empowerment (environment) when I conduct estimation about gender-balanced (pro-environment) firm. Further, in other specification, *Information* is interacted with preference for charity. The function is the form below;

$$Rise \text{ (or } Buy\text{ )}_{it} = b_1 \, Information_t + b_2 \, Information_t * Preference_i$$
$$+ b_3 \, Information_t * Woman_i + m_i + u_{it}, \qquad (1)$$

**IV. Estimation results**

*A. Baseline model*

Tables 2, 4 and 5 present fixed-effects estimates. In baseline model, I see from Panel A of Table 2 that coefficient of *Information* produces the positive sign and statistically

significant at the 1 % level. This implies that giving the peer information raise subjective expected probability about rise of firm stock price regardless of firm types. *Information * Woman*, cross term between *Information* and woman dummy, also show the significant positive sign, with the exception of pro-social firm. This means that woman becomes more likely to raise their expected probability of stock price rising after acquiring the peer information. In Panel B of Table 2, as for subjective willingness to buy firm's stock, similar tendency is observed although *Information * Woman* is not statistically significant in case of Transparent firm.

For closer examination about the Table 2, Table 3 presents the degree of impact of peer information. First columns show the rate of those whose answer is "4" or "5" and so positive for each question before knowing peer information. In cases of pro-environment or pro-social firms, its rates about expected probability of stock price rising was 0.30, which equivalent to rates about willingness to buy. In compared to it, both rates for gender-balanced firm are lower, and these for than those of transparent firms are higher. After knowing these rates, respondents come to have more positive evaluations, with an exception of man's expectation about stock rising for gender-balanced firm. Impact of information is simply calculated based on coefficients shown in Table 2. In each case, coefficient of *"Information"* is divided by rate of those who answered "4" or "5" to the equivalent question in the first wave. To take an example, impact of information on *"Rise pro-environment"* for man is; "0.129" (coefficient of *"Information"*) divided 0.30 (the rate of "4" or "5" to the question). For woman, its impact is: "0.129 (coefficient of *"Information"*) +0.080(coefficient of *"Information* * *Woman"*)" divided by 0.30.

For comparing impact on expectation of stock price rising with that on willingness to buy, *"Buy"* over *"Rise"* is exhibited in the lower parts of Table 3. In most cases, its values are larger than 1, meaning that respondents become more willing to buy than their increase in expectation about stock price rising. This implies that individuals come to put additional values to gender valanced and ESG firms apart from its monetary returns. With the exception of *Transparent firm*, its values using woman sample larger than that using man sample. Interestingly, in case of woman's evaluation about gender-balanced firm, its value is 3.46. which is over two times larger than other values.

Generally, men are more likely to be over-confident and aggressive in stock trading than woman (Barber and Odean, 2001; Cueva et al. 2019). Opposed to it, women are more aggressive to buy the stock of the gender-balanced firm after obtaining peer information. Rate of willingness to buy gender -balanced firm stock is only 0.15, which is smaller value but larger impact for woman than other types of firm. Provision of the information is considered as "nudge" to give woman an incentive to buy. Meanwhile, the rate of *"Buy"* over *"Rise"* is smaller than 1 for woman's evaluation about transparent firm. This indicates that woman is less likely to confident, which is consistent with existing works (Barber and Odean, 2001; Cueva et al. 2019). Well-governed transparent firm is more expected to increase profit. For such typical profit-maximizing firms, woman's cautious characteristics is reflected in the stock trading.

*B. Consideration preference of woman involvement.*

As demonstrated in Figure 2 (a), view about genders varies between man and woman. Therefore, it is plausible that evaluation about *Gender-balanced* firm and the effect of peer information depend on preference for woman involvement. Similarly, there is also difference of preference for environment between genders, which influences the results of evaluation about pro-environment firms. To decompose gender difference and these preferences, Table 4 estimates the specifications by adding cross terms between *Information* and *Prefer woman-involvement,* and *Information* and *Prefer environment.* Unfortunately, questionnaire does not include question related to preference is reasonably related to estimated pro-social and governance of firms. Hence, dependent variables are limited to evaluation about gender balanced and pro-environment firms.

Panel A of Table 4 indicates that interaction term between *Information* and *Prefer woman-involvement* shows the positive sign and statistical significance in columns (1) and (3). Therefore, those who prefer woman-involvement becomes more likely to expect that stock gender balanced firm rise while they are more willing to buy the stock. In my interpretation, before obtaining peer information, preference about woman involvement is unlikely to reflected in individual's evaluation about the gender-balanced firm. However, those who prefer woman involvement express their preference in the evaluation once the peer information is provided. That is, the peer information is considered as "nudge" to reveal individual's preference in behavior. What is more, the significant positive sign of *Information * Woman* continues to be observed although its values of coefficient become smaller than Table 2. Decrease in the values of coefficient can be explained as; Woman tends to prefer woman involvement and so inclusion of *Information*Prefer woman-involvement* partly absorbs influence of *Information * Woman* continues.

When it comes to *Information * Prefer environment,* it shows the positive sign in columns (2) and (4), while being statistically significant only in column (4). That is, effect of environment preference on the expectation about stock price of the pro-environment firm does not change even after acquiring the peer information. However, those who prefer environment becomes more willing to buy stock of the pro-environment firm. Providing peer information leads people who prefer environment to buy the stock even though they do not change their expectation about monetary benefit from the stock.

Turning to Panels B, results using male sample are similar to those using full sample. In Panel C, as for results of female sample, the interaction term shows the positive sign and statistical significance at the 1 % level when willingness to buy the stock is dependent variable. However, neither *Information * Prefer environment* nor *Information * Prefer woman-involvement* are statistically significant regardless of firm's type. That is, peer information does not change woman's expectation about return from the stock, but give a great incentive to reflect her preference in willingness to buy gender balanced and pro-environment firms. Rather than pursuing monetary benefits, they put great importance on role of these firms and so intended to buy its stock to enhance gender balanced and sustainable pro-environment society.

*C. Consideration preference of charity.*

Table 5 reports results to examine how the peer information effects depends on individual's charitable preference. In panel A, I examine how effects of warm-glow and pure altruism are influenced by acquiring peer information after gender difference of peer information effect is controlled. *Information * Donation* captures change of warm-glow effect by the information while *Information * Pure-altruism* captures change of pure-altruistic effect by obtaining the peer information. I observed the significant positive effects of *Information * Donation,* with the exception of columns (4) and (8). Positive effects of warm glow motivation are strengthened by knowing that other's evaluation and willingness to buy gender-balanced and ESG firms. eople with warm-glow motivation are more willing to buy stocks of gender-balanced firm because that they know others are also willing to buy. Interestingly, such effects are not observed in case of transparent firm. This might be because the degree of governance is related to firm's performance in market and monetary benefits for stock holders, but not to other social or environmental improvement.

Meanwhile, surprisingly, *Information * Pure-altruism* does not show statistical significance, with the exception of column (6). Concerning the exceptional case, pure-altruistic individuals become more willing to buy pro-environment firm after acquiring peer information's whereas their expectation about rise of its stock price does not change. This is reasonable that altruistic motivation is unlikely to change in order to pursue self-interest. However, as a whole, effects of pure-altruistic motivation do not change even if people know others' evaluation and willingness. This implies that pure-altruistic intention is firmly self-determined. Results of *Information * Woman* are similar to those in Table 2. Hence, its effects persist even after change of charitable preferences are controlled.

Turning to man sub-sample results, results of Panel B is similar to full-sample results in Panel A. Difference is that *Information * Donation* is not significant about willingness to buy pro-social firm. In addition, *Information * Pure-altruism* shows the significant negative signs when willingness to buy transparent firm. In my interpretation, pure altruistic man reduces their motivation to pursue their self-interest. Switching to woman sample results, Panel C indicates neither *Information * Donation* nor *Information * Pure-altruism* are significant when expectation about stock price rising is examined, with an exception of *Information * Donation* in column (3). However, *Information * Donation* shows the significant positive sign when their willingness to buy stock of gender-balanced, pro-environment, and pro-social firms. Further its absolute values of coefficients are around 0.038 and larger than man results in Panel B. Therefore, in compared with man, woman with warm-glow motivation is more likely to be influenced by other's information when she intends to buy stocks of these firm's which aim to enhance gender balanced sustainable society.

Findings of Table 5 provides evidence that peer information increases warm-glow motivation to buy stock of firms if firms have purpose to promote woman involvement, and sustain environment and society.

All in all, key findings of this study are; (1) provision of peer information leads respondents to expect higher probability of rising of stocks and be more willing to buy it, regardless of types of progressiveness, (2) the effect on willingness to buy is larger than the expected probability of stock price rising. Therefore, apart from returns from stock price, people evaluate the ESG activities and woman empowerment to buy the stock. (3) The effects of providing the information for woman is larger than for man, especially willingness to buy stock of gender-balanced firm. (4) respondents who prefer environment (woman's empowerment) become more willing to buy stock of pro-environment (gender-balanced) firms if they have the information. (5) The effects of providing information are larger for respondents who prefer charitable giving. (6) Only in results of female sample, provision of the information increases the "warm-glow" motivation only for willingness to buy, but not for expectation about rise of stock. However, this is not observed for "transparent firm".

Findings as above implies that provision of peer information is "nudge", especially for woman who are less "confident" than man about stock investment. Further, preference for progressive firm is "revealed" in the investment behaviour in the stock market if the peer information is provided. Especially for woman, provision of the peer information enhances warm-glow incentive to investment for gender-balanced and the ESG firms. The pure-altruism did not enhance investment behaviour is reasonable because investment behaviour is not directly related to help people who encountered the difficulty. It is critical for policy makers to consider interaction "Warm glow" motivation with sharing information about people's view about progressive firms to sustain society.

**VI Conclusion**

Do people consider ESG activities important and so buy the ESG firm stock? However, people possibly buy ESG firm stock to pursue self-interest. Internet experiment enables to scrutinize the motivation of buying stock of the ESG firms. The advantage of this study to use large sample (13,428 observations) and so the peer information is likely to reflect the real situation in the society. There is only two weeks between the first and follow-up surveys. So, it is possible to examine only the effect of provision of the information when other things are equal.

Among ESG, "governance" is different from "environment", "social", and "woman empowerment" because the well-governed transparent firms are expected to improve firm performance to raise the its stock price. People seek their self-interest to put importance on "governance". Comparison between transparent firm and other firms is useful to scrutinize how people are motivated by self-interest.

I found that the peer information raises expected probability of rising of stock price and increase willingness to buy it. Further, the impact of peer information on the willingness to buy is larger than the probability of stock price rising. This implies that peer information leads people to buy ESG firm stock to support the ESG activities. In most cases, effects of the information on woman are larger than on man. However, woman's willingness to buy transparent firm is not larger than man, possibly because transparent firm is expected to improve firm performance to increase stock price. Hence, women are less likely to evaluate self-interest than men. Further, in most cases, individuals who prefer pro-environmental and woman empowerment society, and warm-glow donation become likely to buy the stock if they are informed of peer information. Women who prefer environment and woman empowerment become more willing to buy pro-enviornment and gender-balanced firm's stock even they do not come to expect these firm's stock price to rise. Further, as exceptional case, even after obtaining peer information, people with warm-glow motivation do not change their expectation about stock price of transparent firm and their willingness to buy its stock.

It is follows from findings above that people are motivated to buy the ESG firm stock for supporting sustainable society, but not for self-interest. Provision of peer information is considered as "nudge" to increase individual's confidence about behavior based on their preferences. The "nudge" is more useful for less-confident woman than man.

Figure 1. Comparison of subjective evaluation about firm between genders before experiment.

(a) Gender equalized firm

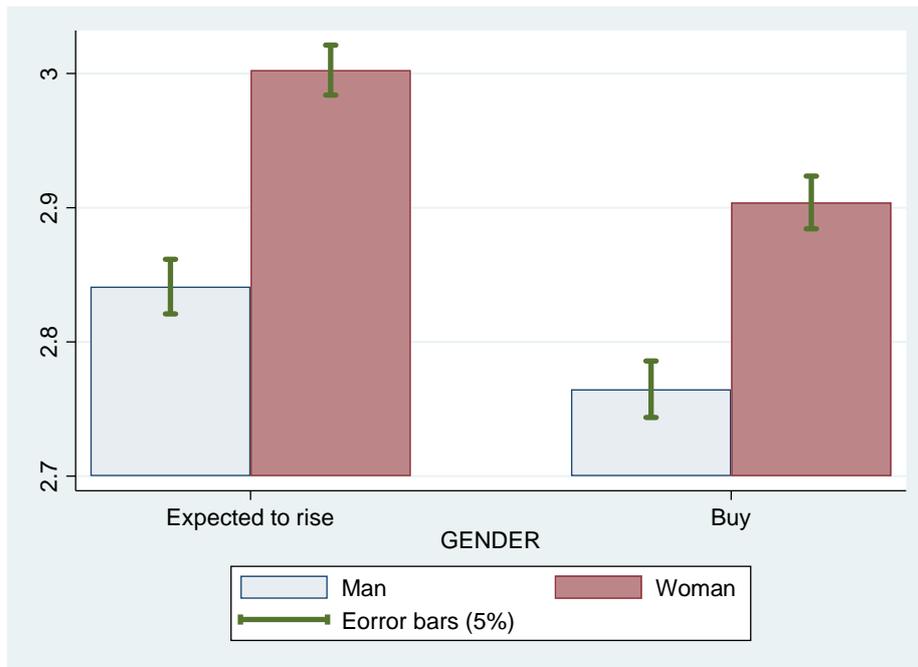

(b) Pro-environment firm

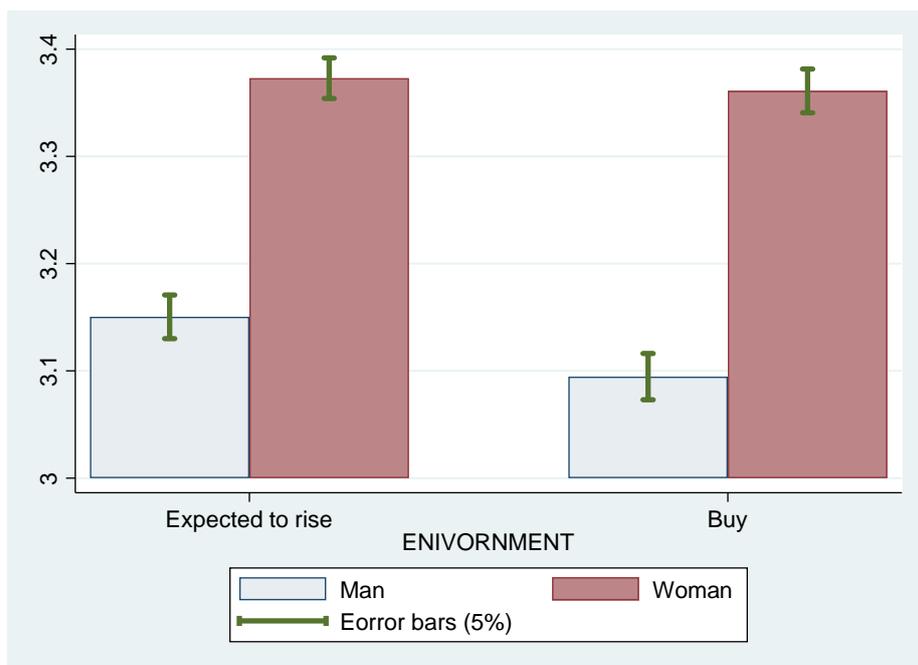

(c) Pro-social firm

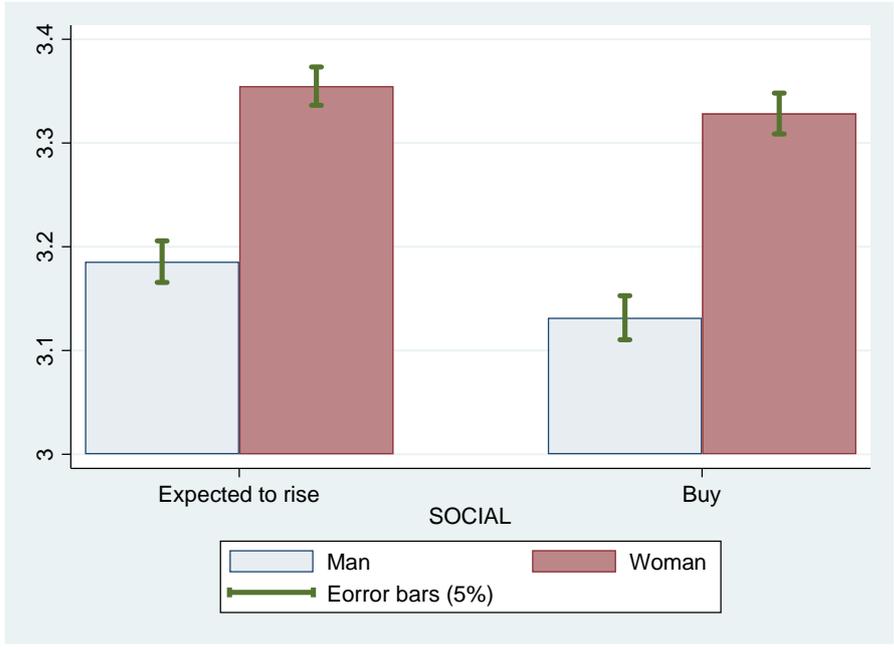

(d) Transparent firms

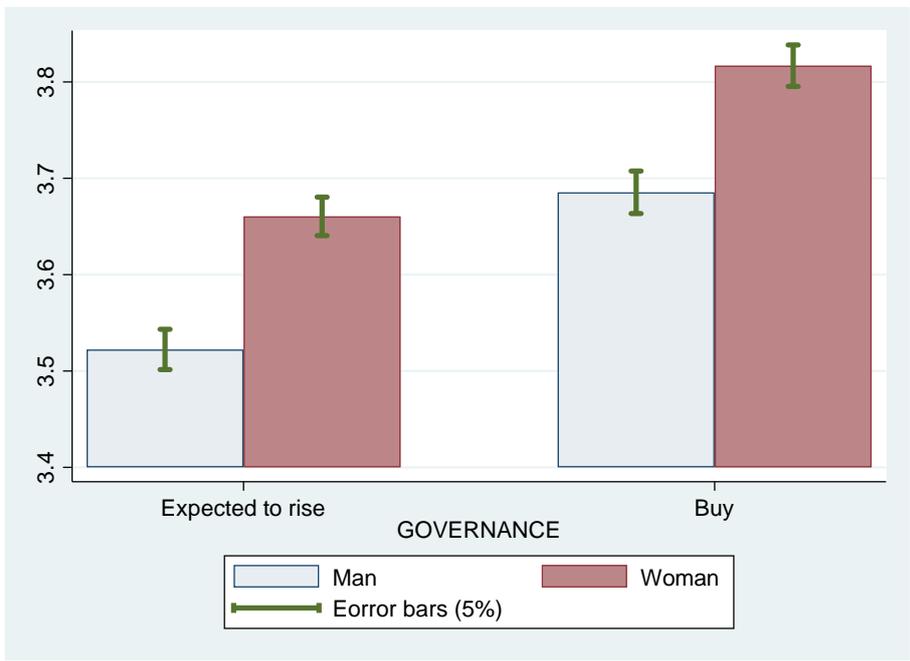

Figure 2 (a). Comparison of preference about firm between genders before experiment.

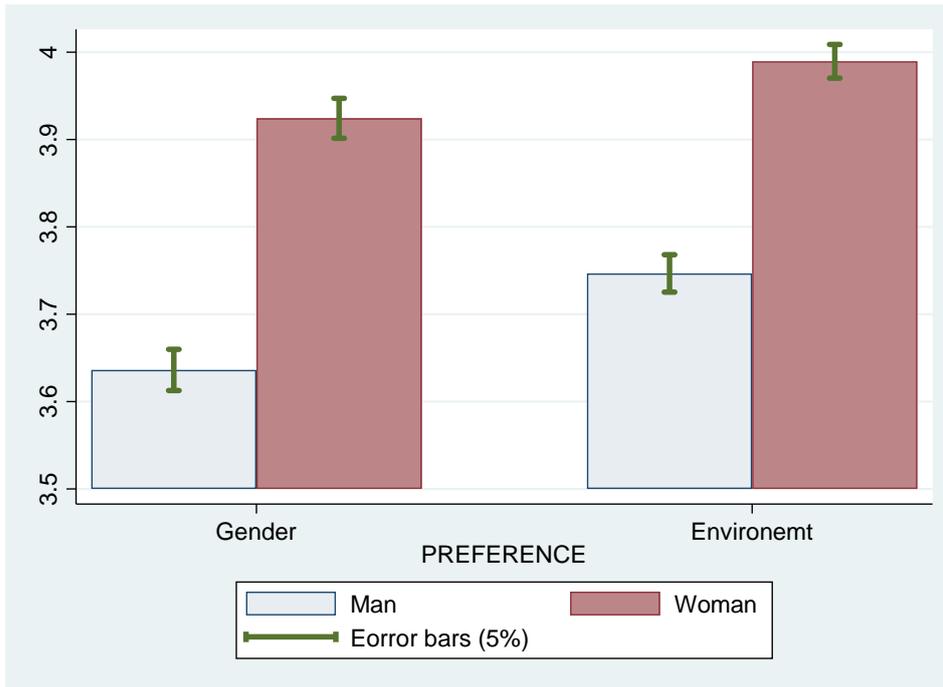

Figure 2(b). Comparison of charitable perception between genders before experiment.

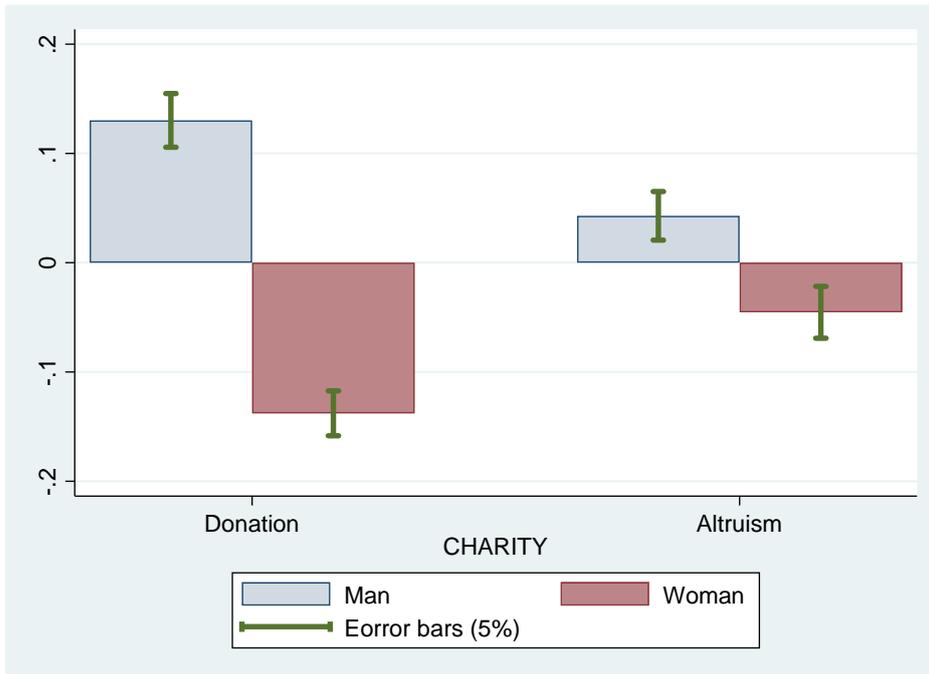

Note: Values of charitable perception are standardized to compare degrees of influence between different variables.

**Table 1**. Definition of variables

| Variables | Definition | Before | After |
|---|---|---|---|
| **Dependent variables** | | | |
| *Rise gender-balance* | *Stock of firm with female board members is expected to rise.* *5 (strongly agree)- 1 (strongly disagree)* | 2.89 | 2.94 |
| *Rise pro-environment* | *Stock of pro-environment firm is expected to rise.* *5 (strongly agree)- 1 (strongly disagree)* | 3.18 | 3.34 |
| *Rise pro-social* | *Stock of pro-social firm is expected to rise.* *5 (strongly agree)- 1 (strongly disagree)* | 3.19 | 3.34 |
| *Rise transparent* | *Stock of transparent firm is expected to rise.* *5 (strongly agree)- 1 (strongly disagree)* | 3.42 | 3.78 |
| *Buy gender-balance* | *Willingness to buy stock of firm with female board members.* *5 (strongly wish to buy)- 1 (no, not at all)* | 2.77 | 2.90 |
| *Buy pro-environment* | *Willingness to buy stock of pro-environment firm* *5 (strongly wish to buy)- 1 (no, not at all)* | 3.12 | 3.34 |
| *Buy pro-social* | *Willingness to buy stock of pro-social firm* *5 (strongly wish to buy)- 1 (no, not at all)* | 3.14 | 3.29 |
| *Buy transparent* | *Willingness to buy stock of transparent firm.* *5 (strongly wish to buy)- 1 (no, not at all)* | 3.54 | 3.98 |
| **Independent variables** | | | |
| *Information* | *It takes 1, if rate of choosing "4 or 5" about subjective view about each type of firm is informed in the second wave, otherwise 0.* | 0 | 1 |
| *Prefer woman-involvement* | *Government should form a society in which women can demonstrate their ability and be actively involved in the work place.* *5 (strongly agree)- 1 (strongly disagree)* | 3.86 | 3.67 |
| *Prefer environment* | *Government should put more emphasis on environment countermeasure..* *5 (strongly agree)- 1 (strongly disagree).* | 3.71 | 4.03 |
| *Donation* | *Assuming that income level of 80% of population is lower than 20% of your income, and amount of tax you paid is directly redistributed to lower income people. What percent of your income can you pay as tax at the maximum?* | 11.8 | |

| | | |
|---|---|---|
| *Pure-altruism* | *On the assumption as above("Donation"), if double of amount of tax you paid is redistributed to lower income people, how do you change amount of tax you can pay?*<br>*Choices; 1(reduce), 2 (same), 3 (increase).* | 1.94 |

Note: Sample is 7855 in each wave. Values in parentheses indicates the rate of those who selected "4" or "5".

**Table 2. Baseline model (Fixed effects estimations)**

**Panel A. Dependent variable: Expected probability about rise of firm's stock price.**

|  | (1) Rise gender-balance | (2) Rise pro-environment | (3) Rise pro-social | (4) Rise transparent | (5) Rise gender-balance | (6) Rise pro-environment | (7) Rise pro-social | (8) Rise transparent |
|---|---|---|---|---|---|---|---|---|
| *Information* | 0.047*** (0.011) | 0.167*** (0.011) | 0.153*** (0.011) | 0.364*** (0.012) | 0.019 (0.015) | 0.129*** (0.016) | 0.136*** (0.017) | 0.334*** (0.017) |
| *Information*Woman* |  |  |  |  | 0.060*** (0.017) | 0.080*** (0.021) | 0.035 (0.020) | 0.063*** (0.021) |
| Within R-square | 0.002 | 0.03 | 0.03 | 0.13 | 0.004 | 0.04 | 0.03 | 0.13 |
| Observations | 13,428 | 13,428 | 13,428 | 13,428 | 13,428 | 13,428 | 13,428 | 13,428 |

**Panel B. Dependent variable: Willingness to buy firm's stock.**

|  | (1) Buy gender-balance | (2) Buy pro-environment | (3) Buy pro-social | (4) Buy transparent | (5) Buy gender-balance | (6) Buy pro-environment | (7) Buy pro-social | (8) Buy transparent |
|---|---|---|---|---|---|---|---|---|
| *Information* | 0.131*** (0.014) | 0.219*** (0.012) | 0.190*** (0.014) | 0.445*** (0.012) | 0.098*** (0.016) | 0.162*** (0.017) | 0.155*** (0.020) | 0.440*** (0.015) |
| *Information *Woman* |  |  |  |  | 0.069*** (0.022) | 0.120*** (0.017) | 0.072*** (0.023) | 0.017 (0.022) |
| Within R-square | 0.02 | 0.05 | 0.04 | 0.17 | 0.02 | 0.06 | 0.04 | 0.17 |
| Observations | 13,428 | 13,428 | 13,428 | 13,428 | 13,428 | 13,428 | 13,428 | 13,428 |

Note: Numbers in parentheses are robust standard errors clustered at residential prefectures. *** indicate significance at the 1% level. Numbers without parentheses are coefficient of each variable.

**Table 3. Evaluation about results of Table 2.**

| Variables | Rate of "4" or "5" in the first wave | Impact of the information. | |
|---|---|---|---|
| | | Man | Woman |
| *Rise gender-balance* | 0.12 | 0 | 0.40 |
| *Rise pro-environment* | 0.30 | 0.43 | 0.70 |
| *Rise pro-social* | 0.30 | 0.45 | 0.45 |
| *Rise transparent* | 0.49 | 0.82 | 0.97 |
| *Buy gender-balance* | 0.15 | 0.82 | 1.39 |
| *Buy pro-environment* | 0.30 | 0.54 | 0.94 |
| *Buy pro-social* | 0.30 | 0.52 | 0.76 |
| *Buy transparent* | 0.41 | 0.89 | 0.93 |
| *Buy gender-balance/ Rise gender-balance* | | ---- | 3.46 |
| *Buy pro-environment / Rise pro-environment* | | 1.26 | 1.35 |
| *Buy pro-social / Rise pro-social* | | 1,14 | 1.67 |
| *Buy transparent / Rise transparent* | | 1.09 | 0.96 |

Note: "Yes" is rate of those who selected "4" or "5". Impact of information is calculated by ("coefficient of *Information*"/rate of "Yes" in the first wave).

**Table 4. Effect of information provision interacted with preference about woman-involvement and environment. (Fixed effects estimations)**

Panel A. Man and Woman

|  | (1) Rise gender-balance | (2) Rise pro-environment | (3) Buy gender-balance | (4) Buy pro-environment |
|---|---|---|---|---|
| *Information* | −0.087 (0.058) | −0.014 (0.063) | −0.163*** (0.055) | −0.245*** (0.065) |
| *Information* * Prefer woman-involvement* | 0.034** (0.014) |  | 0.077*** (0.013) |  |
| *Information* * Prefer environment* |  | 0.025 (0.016) |  | 0.094*** (0.016) |
| *Information* * Woman* | 0.052*** (0.017) | 0.077*** (0.023) | 0.049*** (0.021) | 0.098*** (0.016) |
| Within R-square | 0.02 | 0.05 | 0.04 | 0.08 |
| Observations | 13,428 | 13,428 | 13,428 | 13,428 |

Panel B. Man

|  | (1) Rise gender-balance | (2) Rise pro-environment | (3) Buy gender-balance | (4) Buy pro-environment |
|---|---|---|---|---|
| *Provided information* | −0.160** (0.071) | −0.043 (0.074) | −0.176** (0.074) | −0.236** (0.091) |
| *Information* * Prefer woman-involvement* | 0.055*** (0.018) |  | 0.081*** (0.019) |  |
| *Information* * Prefer environment* |  | 0.034 (0.024) |  | 0.093*** (0.023) |
| Within R-square | 0.02 | 0.04 | 0.03 | 0.05 |
| Observations | 7,048 | 7,048 | 7,048 | 7,048 |

Panel C. Woman

|  | (1) Rise gender-balance | (2) Rise pro-environment | (3) Buy gender-balance | (4) Buy pro-environment |
|---|---|---|---|---|
| *Provided information* | 0.064 (0.073) | 0.106 (0.074) | −0.097 (0.071) | −0.161 (0.102) |
| *Information* * Prefer woman-involvement* | 0.009 (0.018) |  | 0.073*** (0.016) |  |
| *Information* * Prefer environment* |  | 0.013 (0.019) |  | 0.097*** (0.024) |
| Within R-square | 0.03 | 0.08 | 0.05 | 0.12 |
| Observations | 6,380 | 6,380 | 6,380 | 6,380 |

Note: Numbers in parentheses are robust standard errors clustered at residential prefectures. **, *** indicate significance at the 5 and 1% levels, respectively. Numbers without parentheses are coefficient of each variable.

**Table 5. Effect of information provision interacted with preference for charity. (Fixed effects estimations)**

Panel A. Man and Woman

|  | (1) Rise gender-balance | (2) Rise pro-environment | (3) Rise pro-social | (4) Rise transparent | (5) Buy gender-balance | (6) Buy pro-environment | (7) Buy pro-social | (8) Buy transparent |
|---|---|---|---|---|---|---|---|---|
| *Information* | 0.016 (0.016) | 0.124*** (0.016) | 0.131*** (0.017) | 0.333*** (0.016) | 0.093*** (0.016) | 0.156*** (0.017) | 0.152*** (0.012) | 0.435*** (0.015) |
| *Information* Donation* | 0.024** (0.011) | 0.035*** (0.008) | 0.031*** (0.010) | 0.001 (0.011) | 0.037*** (0.009) | 0.034*** (0.011) | 0.024* (0.013) | 0.013 (0.011) |
| *Information* Pure-altruism* | 0.001 (0.009) | 0.008 (0.011) | 0.013 (0.012) | 0.005 (0.008) | 0.010 (0.013) | 0.023*** (0.008) | −0.001 (0.011) | −0.009 (0.009) |
| *Information* Woman* | 0.067*** (0.018) | 0.090*** (0.022) | 0.044** (0.021) | 0.064*** (0.022) | 0.079*** (0.022) | 0.131*** (0.017) | 0.079*** (0.023) | 0.020 (0.022) |
| Within R-square | 0.005 | 0.04 | 0.03 | 0.13 | 0.02 | 0.06 | 0.04 | 0.17 |
| Observations | 13,428 | 13,428 | 13,428 | 13,428 | 13,428 | 13,428 | 13,428 | 13,428 |

Panel B. Man

|  | (1) Rise gender-balance | (2) Rise pro-environment | (3) Rise pro-social | (4) Rise transparent | (5) Buy gender-balance | (6) Buy pro-environment | (7) Buy pro-social | (8) Buy transparent |
|---|---|---|---|---|---|---|---|---|
| *Information* | 0.015 (0.016) | 0.123*** (0.016) | 0.132*** (0.018) | 0.334*** (0.016) | 0.093*** (0.017) | 0.156*** (0.017) | 0.153*** (0.012) | 0.436*** (0.015) |
| *Information* Donation* | 0.025** (0.012) | 0.046*** (0.012) | 0.034** (0.015) | 0.004 (0.012) | 0.034** (0.015) | 0.031** (0.015) | 0.014 (0.018) | 0.019 (0.016) |
| *Information* Pure-altruism* | 0.008 (0.017) | 0.004 (0.020) | 0.014 (0.017) | −0.002 (0.016) | 0.031 (0.020) | 0.034* (0.019) | 0.017 (0.021) | −0.032** (0.015) |
| Within R-square | 0.001 | 0.04 | 0.02 | 0.10 | 0.01 | 0.03 | 0.03 | 0.15 |
| Observations | 7,048 | 7,048 | 7,048 | 7,048 | 7,048 | 7,048 | 7,048 | 7,048 |

Panel C. Woman

|  | (1) Rise gender-balance | (2) Rise pro-environment | (3) Rise pro-social | (4) Rise transparent | (5) Buy gender-balance | (6) Buy pro-environment | (7) Buy pro-social | (8) Buy transparent |
|---|---|---|---|---|---|---|---|---|
| *Information* | 0.081*** (0.014) | 0.212*** (0.017) | 0.175*** (0.013) | 0.398*** (0.018) | 0.172*** (0.018) | 0.287*** (0.001) | 0.232*** (0.015) | 0.455*** (0.016) |
| *Information\* Donation* | 0.022 (0.012) | 0.018 (0.014) | 0.027** (0.013) | −0.002 (0.016) | 0.039** (0.019) | 0.038* (0.019) | 0.037** (0.016) | 0.006 (0.021) |
| *Information\* Pure-altruism* | −0.005 (0.015) | 0.013 (0.015) | 0.013 (0.016) | 0.012 (0.015) | −0.011 (0.018) | 0.012 (0.016) | −0.017 (0.016) | 0.013 (0.018) |
| Within R-square | 0.008 | 0.06 | 0.04 | 0.16 | 0.03 | 0.09 | 0.06 | 0.18 |
| Observations | 6,380 | 6,380 | 6,380 | 6,380 | 6,380 | 6,380 | 6,380 | 6,380 |

Note: Numbers in parentheses are robust standard errors clustered at residential prefectures. ***, ** and * indicate significance at the 1 %, 5 % and 10% levels, respectively. Numbers without parentheses are coefficient of each variable. Values of *Donation* and *Pure-altruism* are standardized to compare degrees of influence between them.

# Appendix:

## Sampling method

We commissioned the Nikkei Research Company to conduct a nationally representative web survey covering all parts of Japan in 2016, 2017 and 2018. We send questionnaire to the same individuals and so constructed panel data. In the survey, we obtained various basic characteristics of respondents such economic condition and demographic information. Web-users are presumably different from non-web users. However, an official survey on information technology indicates that in 2015 nearly 100% of Japanese people in the 20–29, 30–39, and 40–49 age groups are web-users. Even for older age groups, the percentage of web-users is over 90% for people aged 50–59 and 80% for people aged 60–69. Therefore, the sampling method through the Internet is unlikely to suffer bias.

In 2016, the Nikkei Research Company managed to recruit 12,176 people to complete the questionnaire. There are respondents who included in the first survey but did continue to participate in the survey. In 2017, respondents reduced to 9,130. I invited 9,130 subjects to participate in the third survey in 2018. In the last year of the surveys in 2018, we conducted the on-line experiments to examine how peer information changes people's evaluation about ESG firms changed. Actually, in 2018 respondents reduced to 7,855 because some of subjects did not respond. Among 7,855, 6,714 individuals continued to respond to the follow-up survey. In 2018, from the preliminary to the follow-up survey, retention rate is 85.5 % in the experiment.

Figure A1.

(Q18A_33)

先日の調査で、「トップ、幹部に女性がいる企業」の株を買うと答えた人の割合は15%です

Q18A．日本の政治や政策、企業の責任について、あなたのご意見をお伺いします。

| (Q18A_33) | 株を買うとするなら、「トップ、幹部に女性がいる企業」の株を買う。 |
|---|---|

5 　　　同意する
4 　　　やや同意する
3 　　　どちらともいえない
2 　　　あまり同意しない
1 　　　同意しない